\documentstyle[epsfig]{elsart}
\begin{document}
\begin{frontmatter}
\title{Formation of $\eta$-mesic Nuclei Using the Recoilless (d,$^3$He)
Reaction}
\author[Tokyo]{R.S.~Hayano}
\author[Nara]{S.~Hirenzaki} and
\author[Munich]{A.~Gillitzer}
\address[Tokyo]{Department of Physics, University of Tokyo, Hongo,
Bunkyo-ku, Tokyo 113, Japan}
\address[Nara]{Department of Physics, Nara Women's University,
Nara 630, Japan}
\address[Munich]{Physik Department E12, Technische Universit\"at
M\"unchen, D-85748 Garching, Germany}
\date{\today}

\begin{abstract}
We propose to use the recoilless (d,$^3$He) reaction to produce $\eta$-mesic
nuclei.  This reaction has been used to observe deeply bound pionic states
and proven to be powerful recently.  We calculate $\eta$-mesic bound
states in the nucleus using an optical potential and their formation cross
section with the Green function method.  Then, we carefully check the
experimental feasibility. We find that $\eta$-mesic nuclei can be
observed experimentally using the (d,$^3$He) reaction.  We also mention
the possibility to study the formation of $\omega$-mesic nuclei.
\end{abstract}

}
\end{frontmatter}

Bound states of $\eta$ mesons in nuclei ($\eta$-mesic nuclei) are interesting
objects which have not been observed so far experimentally.
The $\eta$-meson is a member of the SU(3) nonet
of pseudoscalar mesons and believed to be one of the Nambu-Goldstone
bosons of spontaneous chiral symmetry breaking. The origin of the mass
of the Nambu-Goldstone bosons is studied theoretically in terms
of a symmetry breaking pattern\cite{klimt}. Since chiral symmetry is
expected
to be partially restored at finite density\cite{review}, it is very interesting
to study the behavior of mesons, especially their masses, in the nucleus.

Since the $\eta$-nucleon scattering
length is dominated by the $s$-wave part due to the strong coupling to
the $N^*$(1535) resonance, the $\eta$-nucleus optical potential has a large
$s$-wave part. Thus by spectroscopic studies of $\eta$-mesic nuclei,
we can expect to obtain precise information on the $s$-wave potential,
which is equivalent to the mass shift of the $\eta$-meson in the nucleus.
We also expect to get new information on the properties of the $N^*$(1535)
resonance in nuclear matter by studying the $\eta$-nucleus optical potential.

The in-medium behavior of $\pi$ and $K$ mesons, which are also
Nambu-Goldstone bosons, has been reasonably well understood from scattering
as well as mesic atom data. In contrast, there is little experimental
information on the in-medium properties of the $\eta$ meson.
Existence of $\eta$-mesic nuclei was suggested theoretically by
Haider and Liu\cite{haider}. They systematically investigated $\eta$-mesic
nuclear states and proposed to use the $(\pi^+,p)$ reaction for their
formation. Experimental attempts to find a bound state in this reaction led
to negative results\cite{chrien}. On the other hand,
the cross section for $\eta$ meson production in d(p,$^3$He)$\eta$
reactions at threshold was found to be large\cite{berger} and was analyzed
in terms of a quasi-bound $\eta-^3$He system\cite{wilkin}.
For the $\eta-^4$He system also the existence of a quasi-bound state
was suggested\cite{willis}.
Recent theoretical work indicated the existence of $\eta$-mesic nuclei,
however their structure is only predicted with large
uncertainty~\cite{chiang,ratityansky}. With the present knowledge,
the existence of $\eta$-mesic nuclei is still controversial.

In this paper we discuss as a new experimental method to produce
$\eta$-mesic nuclei in (d,$^3$He) transfer reactions on light target nuclei,
as recently proposed at the GSI heavy-ion synchrotron SIS~\cite{gsieta}.
The (d,$^3$He) reaction at recoilless kinematics was proven to be a powerful
experimental tool by the discovery of deeply bound pionic atom
formation\cite{hty,gsi160} and to be
very useful to extract the pion properties at finite density\cite{waas,yam98}.
Here we calculate the structure and formation cross section of $\eta$-mesic
nuclear states theoretically and investigate the experimental feasibility.
We should mention here that the same experimental technique
can be used to produce other mesic nuclei. The $\omega$ meson is expected to
be around 16\% lighter at normal nuclear density than in free
space\cite{hatsudalee} and expected to form quasi-bound states.
The $\omega$-mesic nuclear states can also be observed by the (d,$^3$He)
reaction in principle as discussed in Ref.~\cite{gsieta}.


In order to study the structure of $\eta-$mesic nuclei, we used
the first-order in density $\eta$-nucleus optical potential,
\begin{equation}
V_\eta = -\frac{4\pi}{2\mu} \left( 1 + \frac{m_\eta}{M_N}
\right)  a_{\eta N} \rho(r),
\end{equation}
where $a_{\eta N}$ is the $\eta-$nucleon scattering length,
$\mu$ is the reduced mass of the $\eta$ and is $\sim m_\eta$
for heavy nuclei, $M_N$ is the nucleon mass,
and $\rho$ is the nuclear density.

There exist several recent estimates on the
$\eta N$ scattering length:
\begin{eqnarray}
a_{\eta N}&=& [(0.717\pm 0.030) + i (0.263 \pm 0.025)] {\rm fm}
~~~~\cite{batinic},\\
&=& [(0.751 \pm 0.043) + i (0.274 \pm 0.028)] {\rm fm}
~~~~\cite{green},\\
&\approx{}& (0.52 + i 0.25) {\rm fm}
~~~~\cite{willis},\\
&\approx{}& (0.20 + i 0.26) {\rm fm}
~~~~\cite{nkaiser}.
\end{eqnarray}
\noindent
As shown, the first two theoretical estimates
agree fairly well with each other. The third
value was
deduced from an experimental study of $\rm d(p,^3 He) \eta$ and
$\rm d(d,^4 He)\eta$ reactions\cite{willis}.  In all cases,
the $\eta$-nucleus optical potential is expected to be attractive.
For an illustrative purpose, let us
take $\mu = m_\eta = 547$ MeV, $M_N = 939$MeV and
$\rho_0=0.17$fm$^{-3}$
and
$a_{\eta N}= 0.717 + 0.263 ~i {\rm fm}$. We then obtain
$$V(r) = -(86 + 32 i) \rho (r)/\rho_0 {\rm ~MeV},$$
which is indeed strongly attractive.  The imaginary part $W=-\Gamma/2$
is appreciable, but small enough compared with the real part.

With this potential, we calculated the $\eta$-nucleus binding energies and
widths for various nuclei in a conventional way of solving the Klein-Gordon
equation.
The vector part of the potential, which in general must be taken into
account, was ignored in these and the following calculations, based on the
assumption that the $\eta-N$ interaction is dominated by the $s-$wave
component.
A Woods-Saxon form of the nuclear density profile was used, where
nuclear radii and diffuseness were taken to be
$R=1.18 A^{1/3} - 0.48$~fm and $a=0.5$~fm, respectively.
The results  are shown in Table \ref{table:eta1} for the case of $a_{\eta N}
= (0.717 + 0.263 i)$ fm and in Table \ref{table:eta2} for the case of
$a_{\eta N} = (0.20 + 0.26 i)$ fm.
We find that in the former case the half widths are
comparable or smaller than the binding energies and/or level spacings, so
that it is justified to interpret these states as quasi-stable
$\eta-$mesic nuclear bound states.

\begin{figure}[htbp]
\begin{center}
\epsfig{file=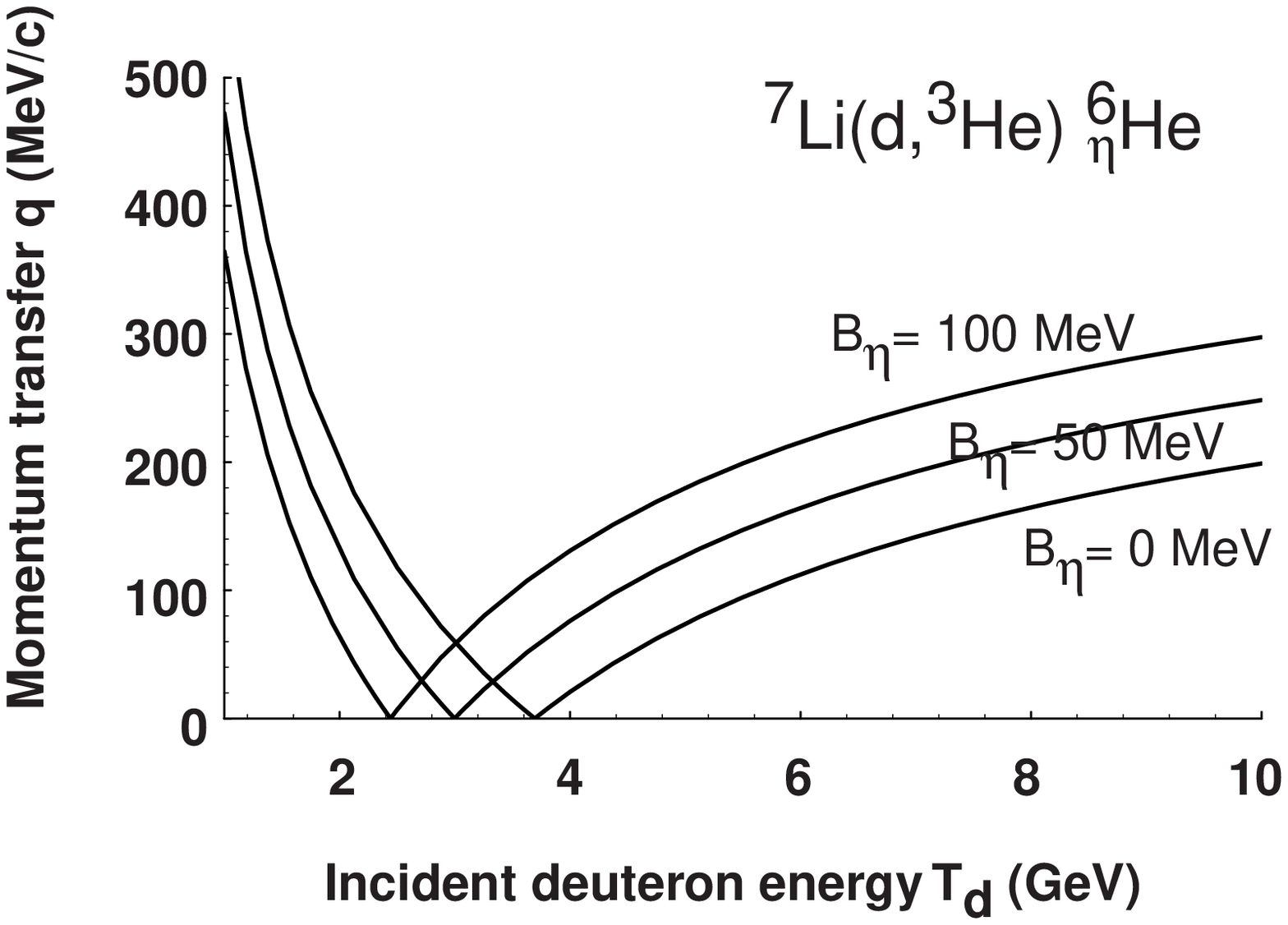,width=10cm}
\end{center}
\caption{\label{fig:qtd}\sl The momentum transfer $q$ vs.
incident deuteron kinetic energy $T_d$ in the $^7$Li(d,$^3$He)$^6_\eta
\sl He$ reaction. The three curves respectively correspond
to $\eta$ binding energies of 100, 50 and 0 MeV, as indicated}
\end{figure}

Similar to the case of deeply-bound pionic atom production, it is possible to
produce $\eta-$mesic nuclei near the recoilless condition using the (d,$^3$He)
reaction on nuclear targets.  This is illustrated
in Fig.~\ref{fig:qtd}, which shows the momentum transfer $q$ vs.
the
incident deuteron kinetic energy $T_d$ for a typical
light target nucleus ($^7$Li in
this case).  The use of recoilless kinematics is essential to suppress the
quasi-free continuum $\eta$ production and to enhance the $\eta-$mesic
nuclear production signal.
The recoil-free $\eta$ condition is satisfied at $T_d \sim 3.6$ GeV and
can be fulfilled at GSI-SIS
where the maximum deuteron kinetic energy $T_d^{max}$ is 4 GeV.

We now estimate the reaction cross section by using the nuclear
response function $S(E)$:

\begin{equation}
 \left( \frac{d^2 \sigma}{d \Omega dE} \right)_{A(d,^3 He)_{\eta}(A-1)} =
 \left( \frac{d \sigma}{d \Omega } \right)_{p(d,^3 He)\eta}
 ^{lab}
 \times
 \sum_{l_{\eta}, j_n, J} S(E)
\end{equation}

\noindent
where
$\left( \frac{d\sigma}{d\Omega}\right)^{lab}_{p(d,^3 He)\eta}$
is the elementary cross section in the laboratory
frame.
A comprehensive and consistent approach to calculate the response
function $S(E)$ for a system with a large imaginary potential was
formulated by Morimatsu and Yazaki\cite{morimatsu}. This method uses
the Green function $G(E; \vec{r}, \vec{r}')$ defined as
\begin{equation}
G(E;\vec{r},\vec{r}') = < p^{-1} | \phi_\eta(\vec{r})
\frac{1}{E-H_\eta+i\epsilon}\phi_\eta^\dagger (\vec{r}') | p^{-1}>,
\end{equation}
where $\phi_\eta^\dagger$ is the $\eta$ creation operator and $|p^{-1}>$ is
a proton hole state.  The Hamiltonian $H_\eta$ contains the $\eta-$nucleus
optical potential.  Since we used energy-independent local potentials in the
present calculation, we can obtain a simple expression for the Green
function as
\begin{eqnarray}
G(E;\vec{r},\vec{r}')&=& \sum_{l_\eta, m_\eta} Y^*_{l_\eta,m_\eta} (\hat{r})
Y_{l_\eta,m_\eta} (\hat{r}') G_{l_\eta} (E;r,r')\\
G_{l_\eta}(E;r,r') &=& -2\mu k u_{l_\eta} (k,r_{<}) v_{l_\eta}^{(+)}
(k,r_{>}) ,
\end{eqnarray}
where $u_{l_\eta}$ and $v_{l_\eta}^{(+)}$ respectively are the radial part
of the regular and outgoing solutions of equation of motion.  Using the
Green function, the response can be calculated as
\begin{equation}
S(E) = - \frac{1}{\pi} Im \sum_{M,m_s} \int d^3r d\sigma d^3r' d\sigma'
f^{\dag}(\vec{r},\sigma) G(E; r, r') f(\vec{r'},\sigma')  .
\end{equation}
We define $f(\vec{r},\sigma)$ as

\begin{equation}
f(\vec{r},\sigma) =  \chi^*_f (\vec{r})
\xi^*_{\frac{1}{2},m_s}(\sigma) [ Y^*_{l_{\eta}} (\hat{r}) \otimes
\psi _{j_p} (\vec{r},\sigma) ]_{JM} \chi_i(\vec{r}),
\end{equation}

\noindent
where $\chi_i$ and $\chi_f$ respectively denote the projectile and
the ejectile distorted waves,  $\psi$ is the proton hole wavefunction
and $\xi$ is the spin wavefunction introduced to count possible spin
directions of the proton in the target nucleus.
The numerical values of $S(E)$ were evaluated by using the eikonal
approximation as in the case of deeply-bound pionic atoms~\cite{toki}.

The elementary cross section for
$\eta$ production which appears in Eq.(6) can be inferred from the energy
dependence of the p(d,$^3$He)$\eta$ cross section measured at
SATURNE\cite{berthet} in the d(p,$^3$He) reaction.
At $T_p=1.75$ GeV (this proton kinetic energy corresponds to the recoilless
$\eta$ production in the p(d,$^3$He)$\eta$ reaction),
the c.m. cross section $(d\sigma /d \Omega)_{cm}$
is 3 nb/sr.  This can be translated to the $d+p$ laboratory-frame cross
section via
\begin{equation}
\frac{d \sigma}{d\Omega}_{lab} = \left(
\frac{p_{lab}(^3{\rm He})}{p_{cm}(^3{\rm He})}\right) ^2
\frac{d\sigma}{d\Omega}_{cm},
\label{eq:boost}
\end{equation}
and the approximate elementary cross section was deduced to be
150~nb/sr.

\begin{figure}
\begin{center}
\begin{minipage}[t]{6.5cm}
\epsfig{file=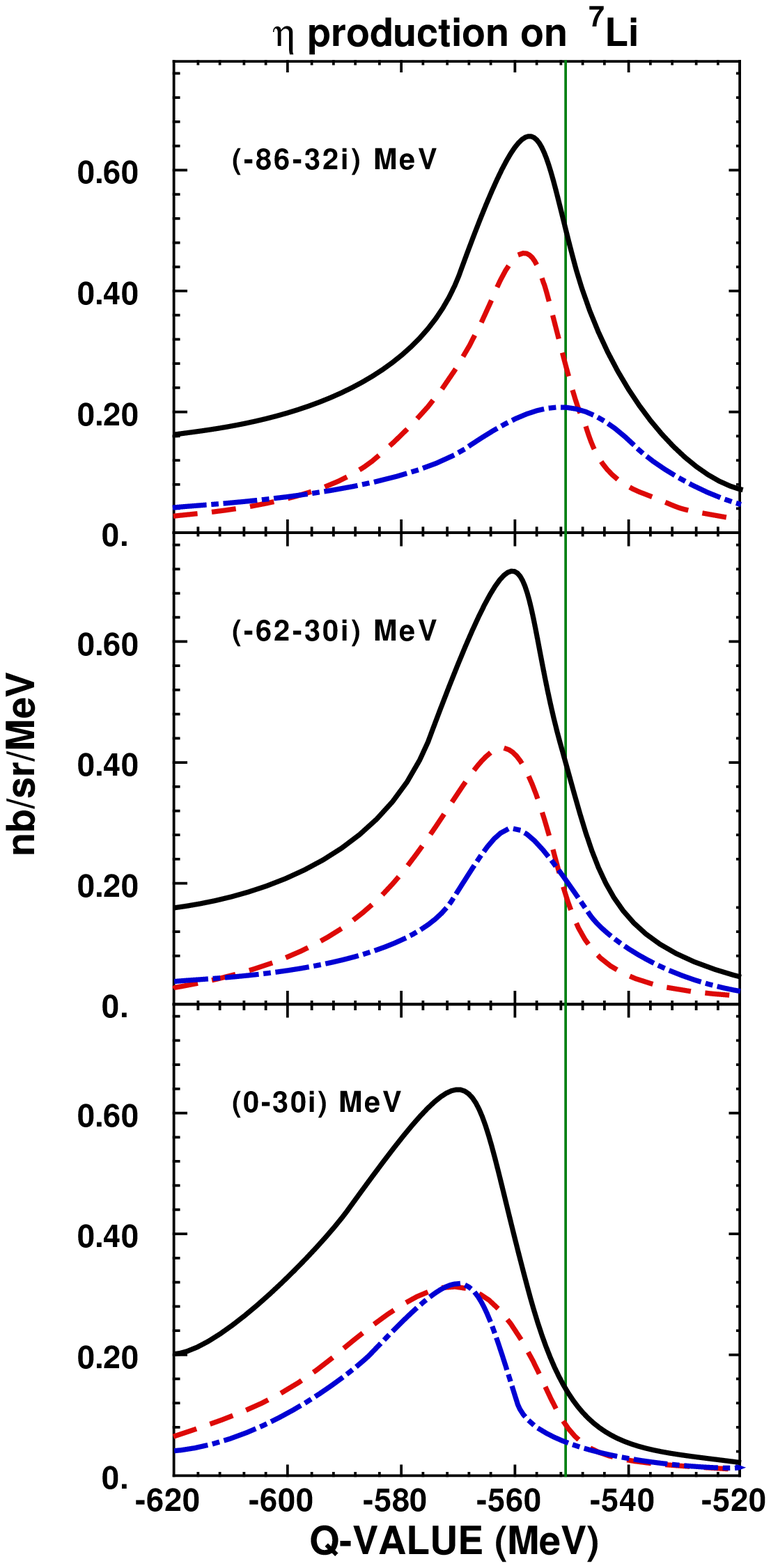,width=6.5cm}
\end{minipage}
\begin{minipage}[t]{7cm}
\hspace*{0.5cm}\epsfig{file=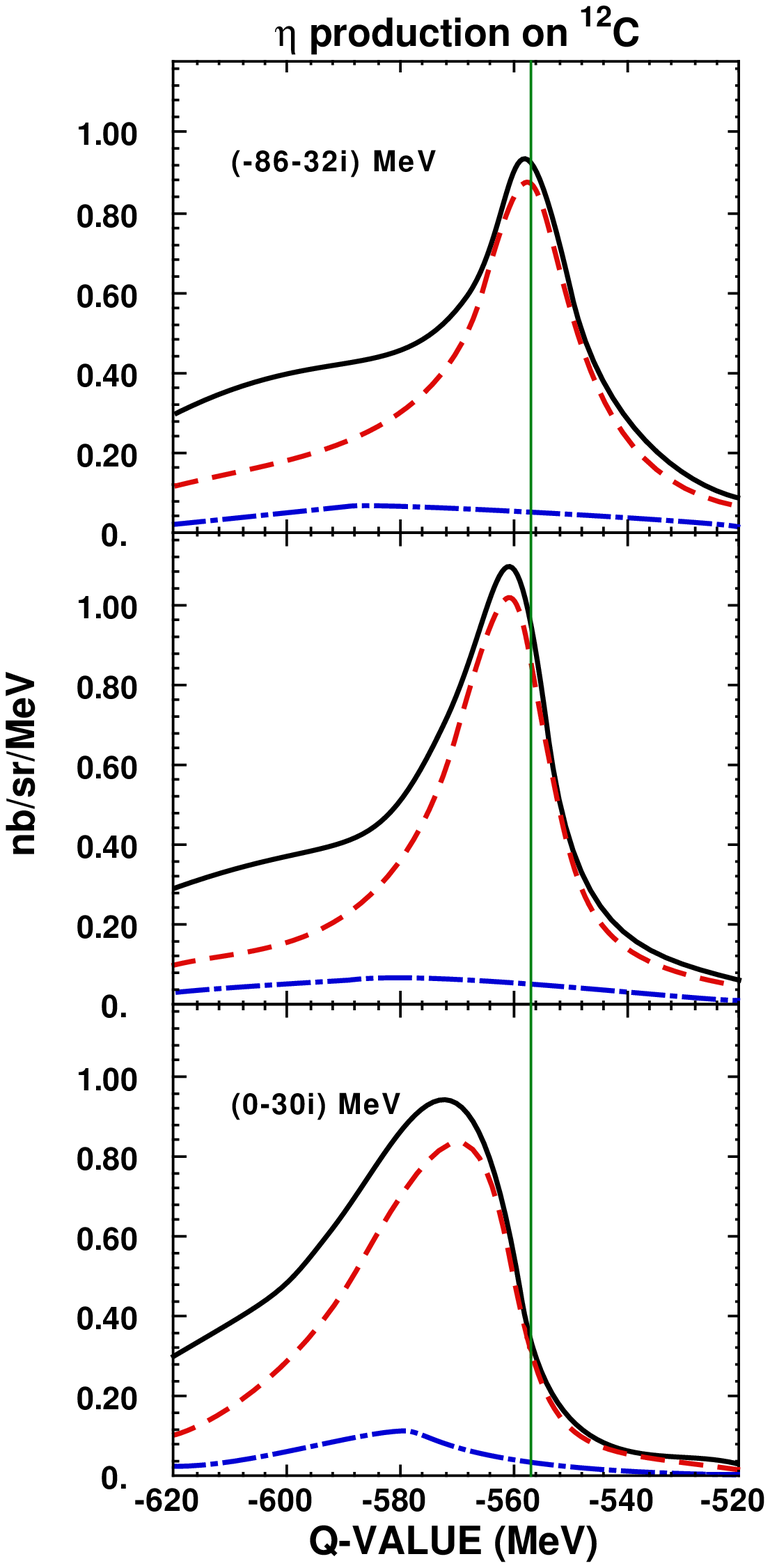,width=6.5cm}
\end{minipage}
\end{center}
\caption{\label{fig:green-eta}\sl The calculated $\eta$ production spectra
for the $^7$Li(d,$^3$He) reaction (left) and for the
$^{12}$C(d,$^3$He) reaction (right) at $T_d=3.5$~GeV,
for three different $\eta$-nucleus
optical potential parameters; (top) $V=-(86+32i)\rho/\rho_0$ MeV, (middle)
$V=-(62+30i)\rho/\rho_0$ MeV, (bottom) $V=-30i \rho/\rho_0$ MeV.  The
vertical lines indicate the $\eta$ production threshold Q-value ($Q_0 =
-552$ MeV for the Li case and -558 MeV for the C case).
In each figure, the contribution from the $(0p_{3/2})^{-1}_{p}\otimes
p_\eta$ is shown in a dashed curve, the $(0s_{1/2})^{-1}_{p}\otimes s_\eta$
contribution is shown in a dash-dotted curve, and the solid curve is the sum of
$\eta$-partial waves up to $l=6$.
The continuum background contributions are estimated to be about 2.7
nb/sr/MeV for the $^7$Li target and 3.4 nb/sr/MeV for the $^{12}$C target
(see text).}
\end{figure}

In Fig.~\ref{fig:green-eta}, we show the calculated
spectra using the Green function method described above. The results are
shown for the $^7$Li target (left panel) and for the $^{12}$C target (right
panel), for different potential parameters.  The top and middle figures
respectively correspond to the $\eta N$ scattering lengths of Eq.(2) ($V(r)
= -(86 + 32 i) \rho(r)/\rho_0$ MeV) and Eq.(4) ($V(r) = -(62+30
i)\rho(r)/\rho_0$ MeV).  The bottom figures are for the potential with no
binding, $V(r) = -30i \rho(r)/\rho_0$ MeV.

In solid lines, the expected double-differential forward ($0^\circ$) cross
sections are shown.  The dashed and dash-dotted lines respectively show the
contributions from the $(p_{3/2})_{p}^{-1}\otimes(2p)_\eta$ and
the $(s_{1/2})_{p}^{-1}\otimes(1s)_{\eta}$ substitutional configurations.
These two configurations contribute dominantly to the Q-value spectra,
although we in fact calculated contributions from other partial waves (up to
$l=6$) and confirmed that there are no significant contributions from
partial waves beyond $l=6$.

The vertical
lines indicate the $\eta$ production thresholds;  for the $^7$Li case, the
threshold is at $Q_0=-552$ MeV while it is at $Q_0=-558$ MeV for the
$^{12}$C case. The $\eta$ binding energy $B_\eta$ can be deduced from the
reaction $Q$ value as (for the sake of simplicity we ignore the nuclear
recoil energy, which is small near the recoilless condition):
\begin{equation}
Q - Q_0 = B_\eta - (S_p(j_p) - S_p(p_{3/2})),
\end{equation}
where $(S_p(j_p)- S_p(p_{3/2}))$ is the proton hole energy measured from the
ground state of the residual nuclei.  Hence, for the $\eta$ states coupled
to the $(s_{1/2})_p^{-1}$ configuration, the
$(\rm{}s_{1/2})_p^{-1}-(\rm{}p_{3/2})_p^{-1}$ energy differences
(14~MeV for $^7$Li and 18~MeV for $^{12}$C) taken from
ref.~\cite{belostotskii} was added when calculating these spectra.

Note that the ground state of the $\eta-$nucleus system for these light
$p-$shell targets would have the $(p_{3/2})_{p}^{-1}\otimes(1s)_\eta$
configuration, but this component does not contribute to the Q-value spectra
near the recoilless condition.  Instead, the dominant contribution comes
from the
$(p_{3/2})_{p}^{-1}\otimes(2p)_\eta$ configuration, and we can determine the
$\eta-$nucleus potential from the location of the $2p$ peak.  This
$(p_{3/2})_{p}^{-1}\otimes(2p)_\eta$ component is more dominant in the
$^{12}$C case because there are four $p_{3/2}$ protons in a $^{12}$C nucleus
as compared to only one in a $^7$Li nucleus.

\begin{figure}
\begin{center}
\epsfig{file=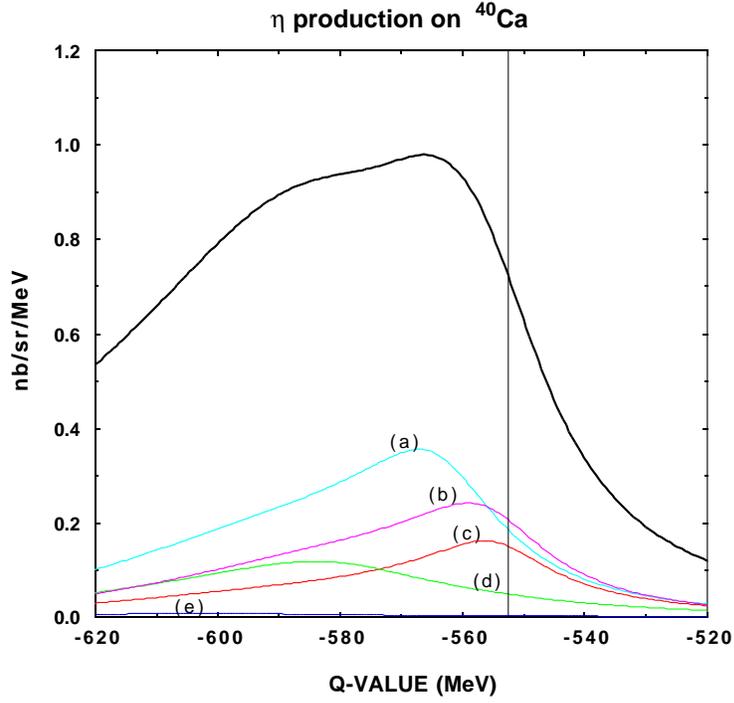,width=10cm}
\end{center}
\caption{\label{fig:ca}\sl The calculated $\eta$
production spectrum for the $^{40}$Ca(d,$^3$He) reaction
at $T_d = 3.5$ GeV, for $V = -(62+30i)\rho/\rho_0$ MeV.  The labelled curves
denote
contributions from the following configurations:
a) $[(1d_{5/2})^{-1}_{p} \otimes d_\eta]$,
b) $[(1d_{3/2})^{-1}_{p} \otimes d_\eta]$,
c) $[(2s)^{-1}_{p} \otimes s_\eta]$,
d) $[(1p)^{-1}_{p} \otimes p_\eta]$ and
e) $(1s)^{-1}_{p} \otimes s_\eta$.
}
\end{figure}

In Fig.\ref{fig:ca},
we show the calculated spectrum for a heavier target, $^{40}$Ca.
The potential strength corresponds to the scattering length in eq. (4).
Due to the recoilless condition, dominant
contributions come from the substitutional states. However, as can be
seen in the figure, several
proton-hole states contribute to a broad maximum in the $Q$ value spectrum
which makes the interpretation difficult. We therefore conclude that
light nuclei like those in the $p$-shell region are most suitable to
search for bound nuclear states of $\eta$ mesons.

In order to generate realistic Q-value spectra we also have to include
background contributions from other reactions, which we will discuss
in the following.

We first note that the background due to $^3$He formation without meson
production (such as due to coalescence) must be negligible.
Composite particle production with a few GeV protons
incident on nuclear targets was studied by
Tokushuku {\it et al.}~\cite{toku} at KEK.  By extrapolating their results on
the deuteron spectra in the
Al(p,d) reaction at $T_p=3~ \rm GeV$ to $p_d \sim 4$~GeV/c,
we found the cross section to be around $10^{-9}\rm{}[nb/(MeV/c)sr]$,
which is negligibly small.

The continuum background for $\eta$ production was estimated by using the
data of Berthet {\it et al.}~\cite{berthet}.
By relating the number of events for $\eta$ production as shown in
figure 1(b) of Ref.~\cite{berthet} to the tabulated c.m. cross section,
we estimated
the c.m.~continuum background level to be $\sim 0.09 $nb/sr/MeV
at $T_p=2$ GeV ($T_d=4$ GeV).
This corresponds to $d^2 \sigma/d E d \Omega_{lab} \sim
4.5$ nb/sr/MeV in the (d,$^3$He) laboratory frame.

The d(p,$^3$He)$\pi^+ \pi^-$ data near the
$\eta$ threshold by Mayer {\it et al.}\cite{mayer} show that the continuum
background due to $\pi^+\pi^-$ production is nearly flat across the $\eta$
production threshold.  We therefore ignored the possible Q-value
dependence of the continuum background.

In order to evaluate the continuum background for the case of
nuclear targets, we need to calculate the distortion effects of the
deuteron and the $^3\rm He$ in the target nucleus.
For this purpose, we summed up the effective numbers for all
final state configurations of proton-hole and mesonic states.  The
calculated effective proton numbers had negligible energy dependence.
We further assumed that the contributions of target protons and neutrons to
the background is identical, and hence multiplied the calculated effective
number by a factor $A/Z$.
This total effective number is expected to be a good estimation of the
distortion effects to the projectile and the ejectile.  And this is also
expected to be consistent with the estimation of signal cross sections.
The effective nucleon number contributing to the background was
calculated, and was used to estimate the constant background level as:
\begin{eqnarray*}
N_{\rm eff} &=& 0.253 \times 7/3 \\
&\rightarrow& \left(\frac{d \sigma}{dEd\Omega}\right)_{\rm background} =
4.5\times 0.253\times 7/3 = 2.7 {\rm nb/sr/MeV} {\rm ~~~ for ^{7}Li}\\
N_{\rm eff}&=&0.373 \times 12/6\\
& \rightarrow& \left(\frac{d\sigma}{dEd\Omega}\right)_{\rm background} = 4.5
\times 0.373\times 12/6 = 3.4 {\rm nb/sr/MeV} {\rm ~~~ for ^{12}C}.
\end{eqnarray*}

\begin{figure}
\begin{center}
\epsfig{file=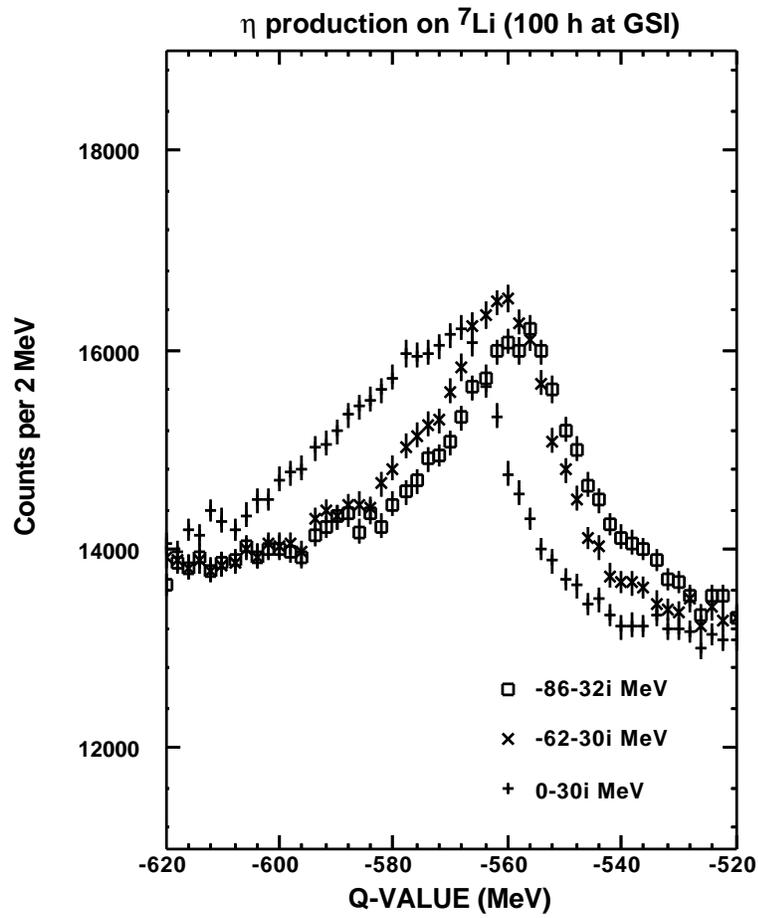,width=10cm}
\end{center}
\caption{\label{fig:expected}\sl Expected Q value spectrum for
the $^7$Li(d,$^3$He) reaction near the $\eta$ production threshold
for 100 hours of running at GSI (see text for assumptions of the
experimental conditions).
}
\end{figure}

In Fig.\ref{fig:expected}, we show expected Q-value spectra for the $^7$Li
case assuming 100 hours of beam time at GSI, using the FRS as $^3$He
spectrometer. As shown, we expect the peaks to be clearly visible above
background, and the spectra are sensitive enough to differentiate between
various $\eta-$nucleus potential parameters.  The experimental setup will be
similar to the one used for the study of deeply-bound pionic atoms.
For the estimate we used a target thickness of $1 \rm g/cm^2$,
a deuteron beam intensity of $3\times 10^{10}$/sec at 3.5~GeV incident
energy, and $\Omega = 2.5 \times 10^{-3}$sr for the FRS acceptance;
all these parameters are achievable at GSI.

\begin{figure}
\begin{center}
\epsfig{file=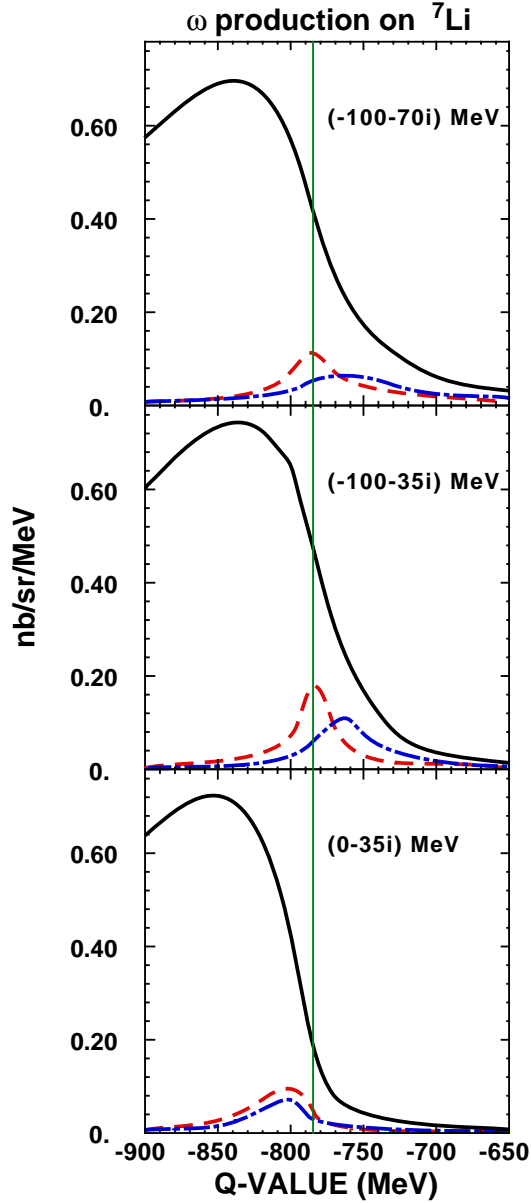,width=7cm}
\end{center}
\caption{\label{fig:omega}\sl The calculated $\omega$
production spectra for the $^7$Li(d,$^3$He) reaction at $T_d=3.8$~GeV,
for three different $\omega$-nucleus
optical potential parameters; (top) $V=-(100+70i)\rho/\rho_0$ MeV, (middle)
$V=-(100+35 i)\rho/\rho_0$ MeV, (bottom) $V=-35 i \rho/\rho_0$ MeV.  The
vertical lines indicate the $\omega$ production threshold of $Q_0 = -787$
MeV.
In each figure, the contribution from the $(0p_{3/2})^{-1}_{p}\otimes
p_\omega$ is shown in a dashed curve, the $(0s_{1/2})^{-1}_{p}\otimes
s_\omega$ contribution is shown in a dash-dotted curve, and the solid curve is
the sum of partial waves up to $l=6$.
The continuum background contributions are estimated to be about 7.7
nb/sr/MeV.}
\end{figure}

Here, we would like to mention that the (d,$^3$He) reaction is also
well-suited for the production of $\omega$-mesic nuclei.
In Fig.\ref{fig:omega}, we show calculated spectra for $\omega$ production
in the (d,$^3$He) reaction on $^7$Li at
$T_d=3.8$ GeV (also possible at GSI), in which we compare three different
$\omega$-nucleus optical potentials.
At this incident energy, however, the recoilless condition
is not satisfied, and we hence
find that the contributions from substitutional states
are not dominant, and that the quasi-free
process makes a large contribution in the unbound region.
Although the identification of bound states appears to be difficult,
the effect of an attractive $\omega$-nucleus potential is noticeable
in the bound region of the $Q$ value spectrum.

\begin{figure}
\begin{center}
\epsfig{file=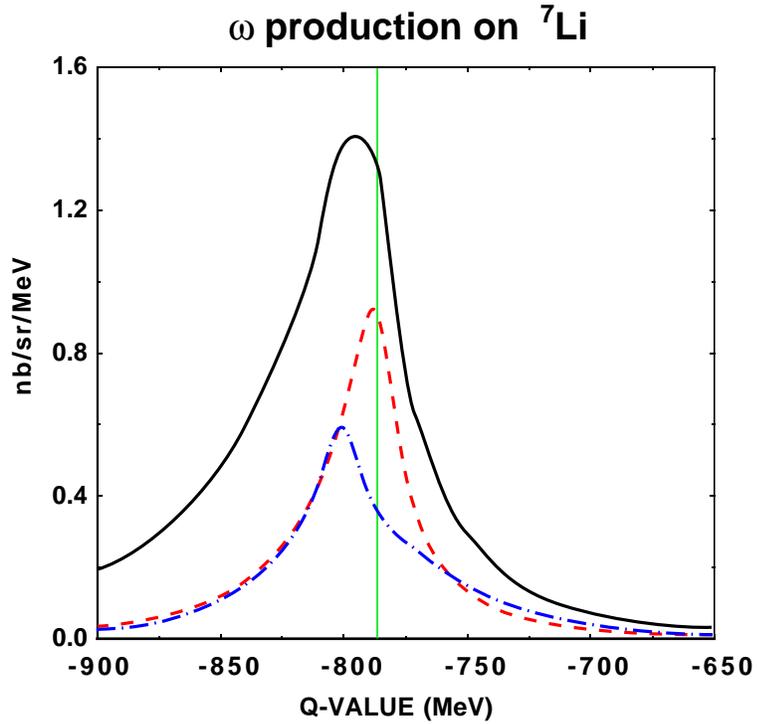,width=10cm}
\end{center}
\caption{\label{fig:omega-recoilless}\sl The calculated $\omega$
production spectrum for the $^7$Li(d,$^3$He) reaction
at $T_d = 10$ GeV. The elementary cross section was assumed to be 450 nb/sr.
The contribution from the $(0p_{3/2})^{-1}_{p}\otimes p_\eta$ is shown in a
dashed curve and the $(0s_{1/2})^{-1}_p \otimes s_\eta$ contribution is
shown in a dash-dotted curve.  The solid curve is the sum of $\omega-$partial
waves up to $l=6$.
}
\end{figure}

In order to see the effect of the recoil free condition for $\omega$
production we show in Fig.\ref{fig:omega-recoilless} the calculated
$Q$ value spectrum at 10~GeV incident deuteron energy.
We assumed the elementary $\omega$ production
cross section to be 450 nb/sr.  In this case, as in the $\eta$-production
spectrum at $T_d = 3.8$~GeV, the dominant contributions result from
substitutional states. With less configurations contributing the
$\omega$-nucleus optical potential may be deduced from the spectral shape
more directly.

One should note that the (d,$^3$He) reaction is the nucleon pickup reaction,
allowing for recoil free $\eta$ and $\omega$ production, with the lightest
projectile and ejectile nuclei.
The (p,d) reaction which may also satisfy the recoil free condition at
appropriate incident energy, however does not allow to separate the
ejectiles
from the beam particles in a magnetic spectrometer due to the same magnetic
rigidity in the case of vanishing momentum transfer.

In conclusion, we find that the proposed recoilless (d,$^3$He) reaction
is a promising tool to study the $\eta-$nucleus system, and we
should be able to determine the $\eta-$nucleus potential (and the possible
$\eta$ mass shift in nuclei) from the Q-value spectra.
In principle, this method can be extended to study the behavior of other
mesons such as $\omega$ in nuclei, although there is at present no facility
in the world where one can study $\omega$ production near the recoilless
condition ($T_d \sim 10$ GeV is required).
The proposed method is complementary to studies of vector mesons in nuclear
matter by analysing their invariant mass spectrum in the dilepton decay
channel, such as $\omega \rightarrow e^+e^-$.
A proposed experimental study of $\eta$ and $\omega$ production in the
(d,$^3$He) reaction at low mometum transfer~\cite{gsieta} was recently
approved at GSI.

The authors would like to thank H.~Toki, T.-S. H. Lee, K. Itahashi, H. Gilg,
F.~Klingl, T. Waas, W. Weise, P. Kienle and T.~Yamazaki for helpful
discussions.
This work is supported in part by the
Grant-in-Aid for Scientific Research, Monbusho, Japan.

\clearpage
\newpage

\clearpage
\newpage

\begin{table}
\caption{\label{table:eta1} \sl A-dependence of the $\eta$-nucleus
binding energies and widths.  We use
$a_{\eta N}=0.717 + 0.263 i$ $[fm]$ as $\eta$-N scattering length.  }
\begin{tabular}{c|cc|cc|cc|cc}
\hline
\hline
A
&\multicolumn{2}{c}{$\ell=0$}&\multicolumn{2}{c}{$\ell=1$}&\multicolumn{2}{c
}{$\ell=2$}
&\multicolumn{2}{c}{$\ell=3$}\\
 & {\footnotesize B.E.(MeV)} &{\footnotesize $\Gamma$ (MeV)} &
{\footnotesize B.E.(MeV)} &
{\footnotesize $\Gamma$ (MeV)} &{\footnotesize B.E.(MeV)} &{\footnotesize
$\Gamma$ (MeV)}
&{\footnotesize B.E.(MeV)} &{\footnotesize $\Gamma$ (MeV)}\\
\hline
6 & 17.4 & 33.5 & & & & & &\\
11 & 35.3 & 48.8 & & & & & &\\
15 & 44.4 & 55.5 & 9.61 & 35.9 & & & &\\
19 & 50.8 & 59.9 & 17.7 & 43.0 & & & &\\
31 & 62.0 & 66.3 & 34.1 & 55.2 & 5.87 & 40.2 & &\\
   & 4.36 & 34.4 & & & & & &\\
39 & 66.4 & 68.2 & 40.8 & 59.1 & 15.0 & 48.0 & &\\
   & 11.8 & 44.5 & & & & & &\\
64 & 74.3 & 71.8 & 53.3 & 63.4 & 31.4 & 58.8 & 10.6 & 52.0 \\
   & 25.8 & 58.2 & & & & & &\\
88 & 77.6 & 73.2 & 61.0 & 66.8 & 40.1 & 59.4 & 21.4 & 60.1\\
   & 33.3 & 56.7 & & & & & &\\
132 & 80.5 & 73.2 & 67.9 & 70.4 & 52.6 & 64.2 & 32.5 & 56.9\\
    & 47.4 & 61.4 & 20.9 & 53.1 & & & &\\
207 & 83.0 & 73.5 & 72.4 & 70.1 & 62.1 & 69.8 & 49.5 & 64.8\\
    & 58.5 & 70.6 & 43.4 & 62.1 & 15.7 & 39.6 & &\\
    & 11.4 & 30.4 & & & & & &\\
\hline
\hline

\end{tabular}
\end{table}

\begin{table}
\caption{\label{table:eta2} \sl A-dependence of the $\eta$-nucleus
binding energies and widths.  We use
$a_{\eta N}=0.20 + 0.26 i$ $[fm]$ as $\eta$-N scattering length.  }
\begin{tabular}{c|cc|cc|cc|cc}
\hline
\hline
A
&\multicolumn{2}{c}{$\ell=0$}&\multicolumn{2}{c}{$\ell=1$}&\multicolumn{2}{c
}{$\ell=2$}
&\multicolumn{2}{c}{$\ell=3$}\\
 & {\footnotesize B.E.(MeV)} &{\footnotesize $\Gamma$ (MeV)} &
{\footnotesize B.E.(MeV)} &
{\footnotesize $\Gamma$ (MeV)} &{\footnotesize B.E.(MeV)} &{\footnotesize
$\Gamma$ (MeV)}
&{\footnotesize B.E.(MeV)} &{\footnotesize $\Gamma$ (MeV)}\\
\hline
31 & & & & & & & &\\
39 & 5.25 & 51.9 & & & & & &\\
64 & 9.41 & 57.2 & & & & & &\\
88 & 11.7 & 59.0 & & & & & &\\
132 & 14.2 & 60.6 &  & & & & &\\
207 & 16.2 & 61.8 & 9.09 & 57.2 & & & & \\
\hline
\hline

\end{tabular}
\end{table}

\end{document}